\documentclass[prd,superscriptaddress,twocolumn,nopreprintnumbers,floatfix,nofootinbib]{revtex4}

\usepackage[nodayofweek]{datetime}
\usepackage{color,float}
\usepackage{eso-pic}
\usepackage{subfigure}
\usepackage{graphicx}
\usepackage{mathrsfs}
\usepackage{hyperref}

\begin{document}
\title{A fifty-fold improvement of thermal noise limited inertial sensitivity by operating at cryogenic temperatures}
\author{J.V. van Heijningen$^{ \dagger,}$}
\affiliation{ARC Centre of Excellence for Gravitational Wave Discovery OzGrav}
\affiliation{The University of Western Australia, 35 Stirling Hwy, Crawley WA 6009, Australia}
$^{\dagger}$joris.vanheijningen@uwa.edu.au
\date{\today}

\begin{abstract}
A vacuum compatible cryogenic accelerometer is presented which will reach $<0.5$\,p$g$ Hz$^{-1/2}$ sensitivity from 1\,mHz to 10\,Hz with a maximum sensitivity of 10\,f$g$\,Hz$^{-1/2}$ around 10\,Hz. This figure can be translated to a displacement sensitivity $<2$\,fm\,Hz$^{-1/2}$ between 2 - 100\,Hz. This will supersede the best obtained sensitivity of any motion sensor by more than three orders of magnitude at 1\,Hz. The improvement is of interest to the fields of gravitational wave instrumentation, geophysics, accelerator physics and gravitation. In current particle accelerators and proposed future gravitational wave detectors $<$\,10\,K cryogenics are applied to the test masses in order to reduce thermal noise. This concept can benefit from the already present superconducting regime temperatures and reach a $>$\,$10^5$ signal-to-noise ratio of all terrestrial seismic spectra. The sensor may be used for control of beam-focusing cryogenic electromagnets in particle accelerators, cryogenic inertial sensing for future gravitational wave detectors and other fields.
\end{abstract}

\maketitle

\section*{Introduction}
Since 1962, gravitational wave scientists have been pursuing an interferometric approach to probe space-time curvature ripples\,\cite{IFOsuggest}. With first the detection of gravitational waves (GWs)\,\cite{GWfirst}, the most precise distance measurement ever was made. The first coincidental measurement of GWs with electromagnetic counterparts, GW170817, from a binary neutron star merger\,\cite{MMgw, MMall} has provided a firm basis for the newly founded field of multi-messenger gravitational wave astronomy and an independent confirmation the gravitational wave detector measurements. In future, low frequency GW detections will give access to heavy mass black hole insprial signals.

All these monumental measurements would not have been possible without decoupling the test masses of the detectors from the Earth's ever-present motion. The \textit{seismic wall}, after the appropriate vibration isolation, is typically limiting below 10\,Hz. Many of the world's most precise commercial sensors were used in LIGO\,\cite{LIGO} and Virgo\,\cite{iVirgo} and continue to be used in Advanced LIGO (aLIGO)\,\cite{aLIGO}, Advanced Virgo\,\cite{AdVirgo} and KAGRA\,\cite{KAGRA}. Some custom made sensors were also researched and developed, such as the LVDT\,\cite{LVDT} or the OSEM\,\cite{OSEM} for differential sensing. For the angular degree of freedom, the Beam Rotation Sensor (BRS)\,\cite{BRS} and A Low Frequency Rotational Accelerometer (ALFRA)\,\cite{ALFRA} have been developed. Currently the Precision Laser Inclinometer (PLI)\,\cite{PLI} is being installed in Advanced Virgo.

\begin{figure}[h]
\centering
\includegraphics[width=0.48\textwidth]{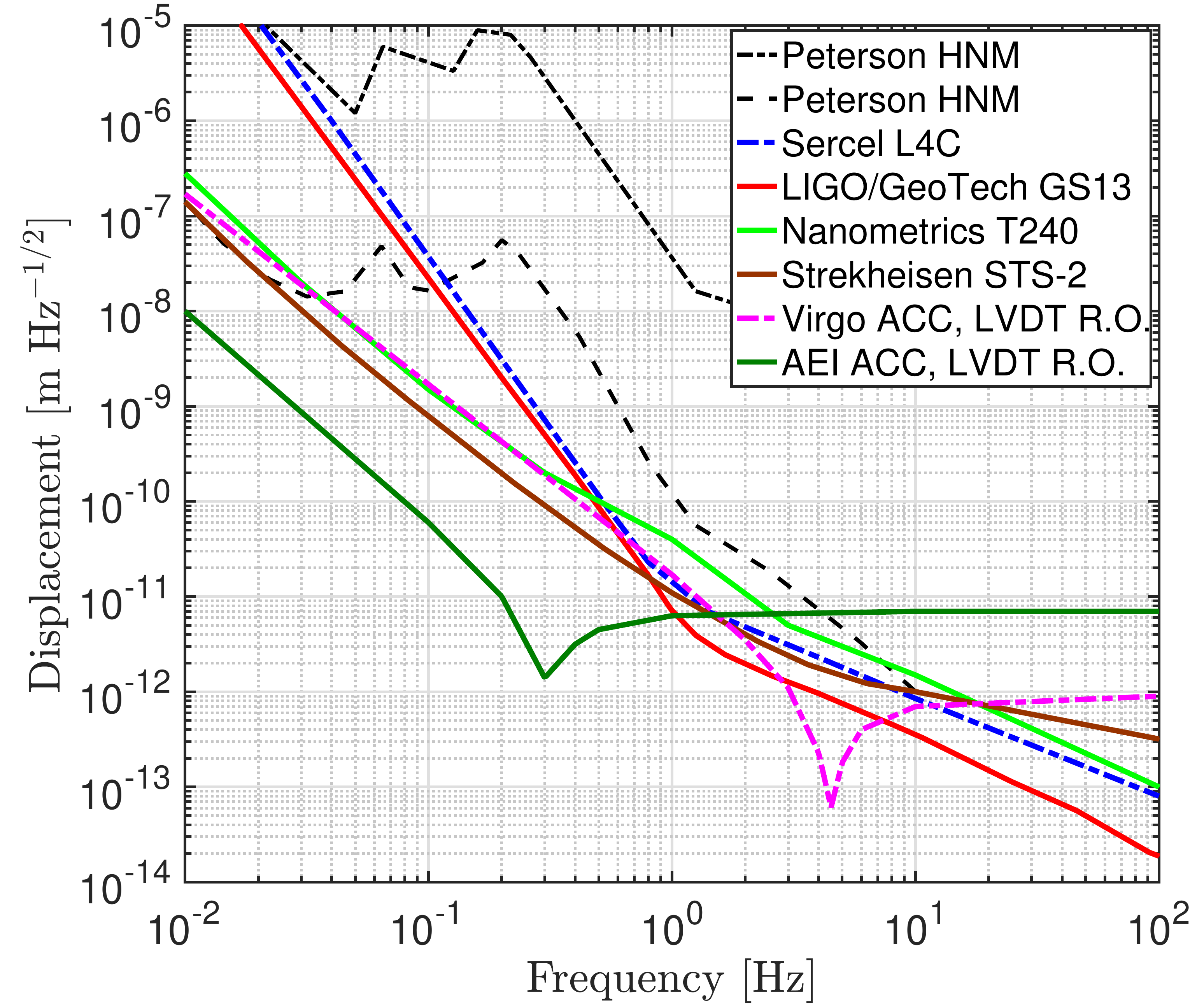}
\caption[Measured or specified displacement sensitivity for several inertial sensors]{Measured or specified displacement sensitivity for inertial sensors used in geophysical and gravitational wave experiments. The Peterson high and low noise models (HNM/LNM) data are from ref.\,\cite{Peterson}}
\label{fig:senscomp}
\end{figure}

The inertial sensors used in the field of GW instrumentation are mostly commercial, \textit{e.g.} the Sercel L4C\,\cite{L4C} or the Geotech GS13\,\cite{GS13Spec}, but some custom built accelerometers have been developed for use in the Virgo superattenuator\,\cite{SA}. In Fig.~\ref{fig:senscomp} the commonly used inertial sensors are compared. Note that low frequency performance is ignoring any angular-to-horizontal coupling. In practice, a matching tiltmeter with sufficient sensitivity to measure angular motion to correct for this inevitable coupling is needed.

Many sensor performances displayed in Fig.~\ref{fig:senscomp} can be used to measure almost all locations on Earth with reasonable signal-to-noise ratio (SNR) as the sensitivy is below the Peterson Low Noise Model. Some are sufficiently sensitive at high frequency to actively damp an inertial platform as used in aLIGO suspensions. Outside gravitational wave physics, geophysics, accelerator physics and gravitation can benefit from even better performance.  

A superconducting gravimeter has been presented and used in the past\,\cite{Paik}, where a Superconducting QUantum Interference Device (SQUID) was used for its readout. Acceleration sensitivities of about 10$^{-10}$\,m\,s$^{-2}$\,Hz$^{-1/2}$ were demonstrated. Additionally, Microelectromechanical system (MEMS) accelerometers have entered the stage for gravimeters\,\cite{wee-g} and accelerometers\,\cite{MEMS}. Both options reach n$g$\,Hz$^{-1/2}$ sensitivities; the former even reaches down to 10$^{-6}$ Hz. 

Recently, an interferometric readout\,\cite{Gray} has been combined with a monolithic accelerometer\,\cite{2006} at Nikhef. A prototype was made and measurements were performed\,\cite{IFOsens1,IFOsens2}. Bench motion of 8$\cdot 10^{-15}$\,m\,Hz$^{-1/2}$ from 30\,Hz onward was measured limited by the sensor self-noise. Continued development to reach the modelled sensor self-noise of 3$\cdot\,10^{-15}$\,m\,Hz$^{-1/2}$ from 10\,Hz is ongoing.

Here, a concept for a sensor is presented that will enhance inertial sensitivity by at least two orders of magnitude between 10\,mHz - 100\,Hz compared to the state of the art. It uses the superconducting characteristic of the proposed mechanics material Niobium to decimate the effect of Eddy current damping in the coil magnet actuator. Section~\ref{des} will discuss the proposed design.The effect the superconductive state has on the accelerometer mechanics on Eddy current damping is discussed in section~\ref{sc}, which will result in an modelled noise budget in section~\ref{nb}. Possible applications are discussed in section~\ref{pa} after which a conclusion is provided.

\section{Proposed design and material choice}\label{des}

\begin{figure}[h]
\centering
\includegraphics[width=0.45\textwidth]{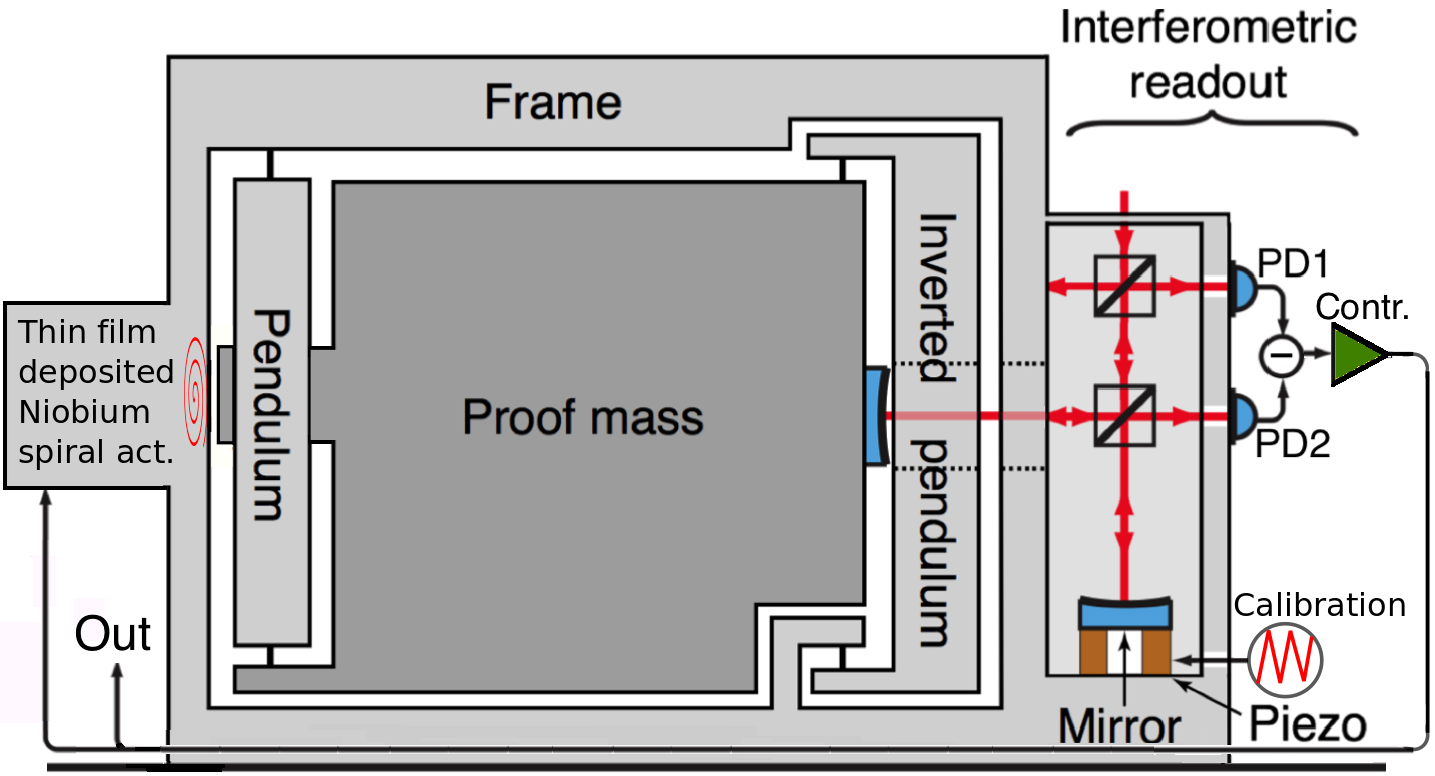}
\caption{Niobium monolithic accelerometer with interferometric optical readout. The position of the proof mass is probed by an interferometer with a differential readout. A piezo actuated mirror is used for calibration outside regular operation. The difference between the two interferometer output signals is kept null by a feedback loop. The feedback loop uses a thin film deposited Niobium spiral as an actuator. It keeps the mass at a fixed position with respect to the frame and the signal it needs to do that can be used as sensor output. This design is a combination of accelerometer mechanics\,\cite{Bertolini} and an interferometeric readout\,\cite{Gray}.}
\label{fig:SetupStart}
\end{figure}

The low frequency part of the Ref.\,\cite{IFOsens2} noise budget and possibly also the measurement is obscured by suspension thermal noise. A disappointing quality factor of 40 was determined for the mechanics of the accelerometer\,\cite{MyThesis}. Shot noise was the dominant noise force from about 10\,Hz in the designed noise budget. The design is shown in figure~\ref{fig:SetupStart}.

An interferometric readout with a voice coil actuator to keep the proof mass mirror position in the linear regime of the interferometer fringe can prove a superior sensing solution with a relatively high dynamic range. The dynamic range is set by the quality of the readout and control electronics and can be as high as eight orders of magnitude. 

\section{Changes to superconductive mechanics and actuators}\label{sc}

A switch to a material that becomes superconducting at cryogenic temperatures could decrease the effect of Eddy current damping. Niobium seems to be the most logical choice as it has a transition temperature at 9.2\,K, high strength and high intrinsic quality factor. Niobium has been used for bar detectors\,\cite{NIOBE} and suspensions for gravitational wave detectors\,\cite{Lee} mostly because of these favourable characteristics. 

One of London's equations is a result of manipulating Ampere's law and governs the (highly reduced) penetration depth of the magnetic field in a superconducting material as\,\cite{LNDeq}
\begin{equation}
\nabla^{2} \mathbf{B}=\frac{1}{\lambda^{2}} \mathbf{B},~\rm{with} \quad \lambda \equiv \sqrt{\frac{m}{\mu_{0} n_{\rm{pl}} e^{2}}}.
\end{equation}
Here, $\bf{B}$ denotes the magnetic field within the superconductor, $\lambda$ the London penetration depth, $m$ the mass of the charge carrier, $\mu_0$ the magnetic permeability in vacuum, $n_{\rm{pl}}$ the planar density and $\rm{e}$ the charge of the carrier.

Niobium has a BCC lattice, therefore $n_{\rm{pl}} = 5 \cdot 10^{18} \rm{cm}^2$ and, using electron characteristics for the charge carrier, $\lambda$ is determined to be about 3\,$\mu$m. This means that the magnetic field decays exponentially to a negligible value within 20\,micron and, since currents are practically loss-less in a superconductor, Eddy current damping is therefore assumed not to be dominant over structural damping in the following discussion.

The $Q$ of the Niobium mechanics may be assumed to be about $10^4$ in the cold state\,\cite{PCKann}. The actuator will be conceived as a thin film superconducting \textit{coil}, similar to the designs used for cryogenic bar detector readout schemes. Thin film deposited Niobium spiral actuators are used.\,\cite{PresKann}. The actuator design will not affect the overall mechanical $Q$ as its (reduced) effect is summarized by stating this electromechanically coupled damping channel has a $Q~>~10^5$\,\cite{PCKann}. Both these considerations validate the assumption made below on the $Q$ and its subsequent fifty-fold reduction of the thermal noise.

The use of the spiral actuator that will generate a magnetic field pressure on the extrusion shown in Fig~\ref{fig:SetupStart} as that volume will portray the Meissner effect. The \textit{push only} actuator will act as a spring which could possibly spoil the sensor performance by injecting frame motion in the inertial mass. The magnetic pressure is given by $p_{\rm{mgn}} = B/(2 \mu_0)$, where $B$ is the magnetic field strength at the extrusion surface. Assuming an area of 1\,cm$^2$ of the spiral actuator, the actuator force is
\begin{equation}\label{eq:ActF}
    F_B = 5\cdot10^{-5}\frac{B^2}{\mu_0} \approx 40 B^2.
\end{equation}

The standing force, as it will generate a magnetic field that is uniform on a small scale, will not result in an actuator noise. The application of the Biot-Savart law on the center-line of a current loop involves integrating the $z$-component, where $z$ is the axis normal to the loop. As an example design, the actuator is modelled as 10 loops in a 1\,cm$^2$ area with radius $R$ between 0.05\,cm\,$\leq R \leq$\,0.5\,cm, and this yields
\begin{equation}
    \frac{B_z}{I} = \frac{\mu_0}{4 \pi} \frac{2 \pi R^2}{{(z^2+R^2)^{3/2}}} \approx 1.8\cdot10^{-3}~\rm{T/A},
\end{equation}
where $I$ is the supplied current to the actuator. Substituting this result in Eq.~\ref{eq:ActF} and considering a typical actuator current of 10\,mA yields a force associated with the supplied $B$ field of $F_B = $12.96\,nN when assuming a 0.1\,mm actuator gap. The Peterson high noise model peaks around 1.5\,$\mu g/ \rm{Hz}^{-{1/2}}$, which would require 1 A current supply to the modelled actuator design. Depending on the application the acceleration needed to keep the interfereometric readout in the linear part of the fringe, the same actuator design may not need such amount of current.

Assuming the proof mass moves with an amplitude of 1\,micron during usual operation, the stiffness of this spring (assuming a roughly constant $F$ when supplying the said 10~mA) is about $k_{\rm{act}}=0.013~$N/m. The spring constant of a Watt's linkage with a 1 kg proof mass tuned to 0.4~Hz is about $k_{\rm{mech}}=6.31~$N/m which is almost a factor of 500 higher than $k_{\rm{act}}$. The actuator's impact on the overall stiffness is therefore negligible.

Further reduction of $k_{\rm{act}}$ is possible by
\begin{itemize}
    \item a more homogeneous magnetic actuator, \textit{e.g.} using thin film Niobium loops deposited on silicon  wafers stacked to make a coil. Niobium wires are not chosen here as historic practical measurement\,\cite{DBcomm} of surface loss of the fabricated wire surfaces are too high - thin films can be produced with much less surface loss. Removing the centre of these loops allows for flag insertion allowing probing of the most homogeneous part of the $B$ field;
    \item applying a wedge to the extrusion and tuning the suspension points of the Watt's linkage such that horizontal motion couples to vertical motion\,\cite{EHcomm}. This way, the average gap between actuator spiral and extrusion  surface is such that the overall stiffness can be arbitrary low.
\end{itemize}
All above considerations prove the superconducting spiral actuator, even without much changes from earlier design principles, is viable for its purpose. The calculated $F_B$ also suggests that typical stray AC magnetic fields are not worrisome as a potential noise source. 

The Earth's typical magnetic field has a magnitude around 50 $\mu$T and has varied from 56 to 52.5\,$\mu$T from 1970 to 2012\,\cite{EarthMagn}. The variations on the 1\,mHz scale are many orders of magnitude smaller. Therefore, the Earth magnetic field can be omitted from stray field issue analysis. Careful design of surrounding magnetic sourcing machinery or actuators must be observed not to spoil the sensor performance.

In any physics experiment, stray magnetic fields generated by some device could interfere with the operation of another device. For this accelerometer this interference can occur in two distinct ways. First, the magnetic field can couple to the proof mass and introduce a acceleration noise in a mechanical sense. This could be mitigated by use of a solid box of superconducting material around the full accelerometer. The Meissner effect of that box will act as a Faraday's cage for magnetic fields. Lead is easily machined and weldable and has appropriate superconducting characteristics and can be used for this.

Second, the PDs and subsequent readout electronics might be affected by strong magnetic fields. To solve this, already research towards fully separating the optical readout and its conversion to electronic signals. More results are found in ref.\,\cite{mythesis}, but the effort can be summarised by stating a pm\,Hz$^{-1/2}$ sensitivity was obtained using optical fiber. An in-fiber scheme using fiber splitters, circulators and fiber PDs was used to show proof-of-principle for the room temperature sensor in context of its deployment in the proposed CLiC linear collider at CERN. Linearity in the in-house made piezo fiber stretcher actuators was shown and a solid comparison to a Sercel L4C geophone was presented.

\section{Noise budget of readout and mechanics}\label{nb}

\begin{figure}[h]
\centering
	\subfigure[~]{
	\includegraphics[width=0.47\textwidth]{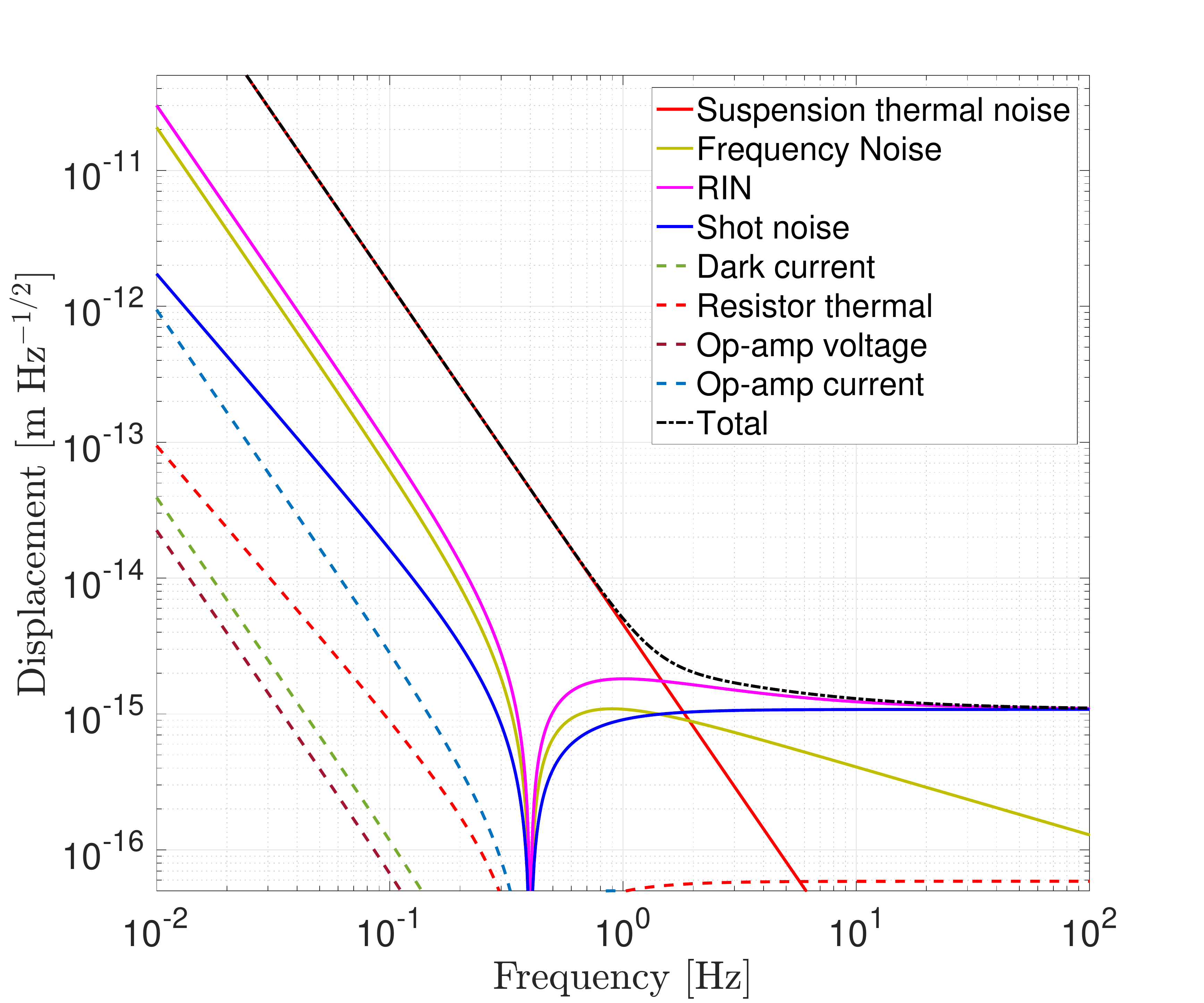}
    \label{sfig:NBmain_d}}
	\subfigure[~]{
	\includegraphics[width=0.47\textwidth]{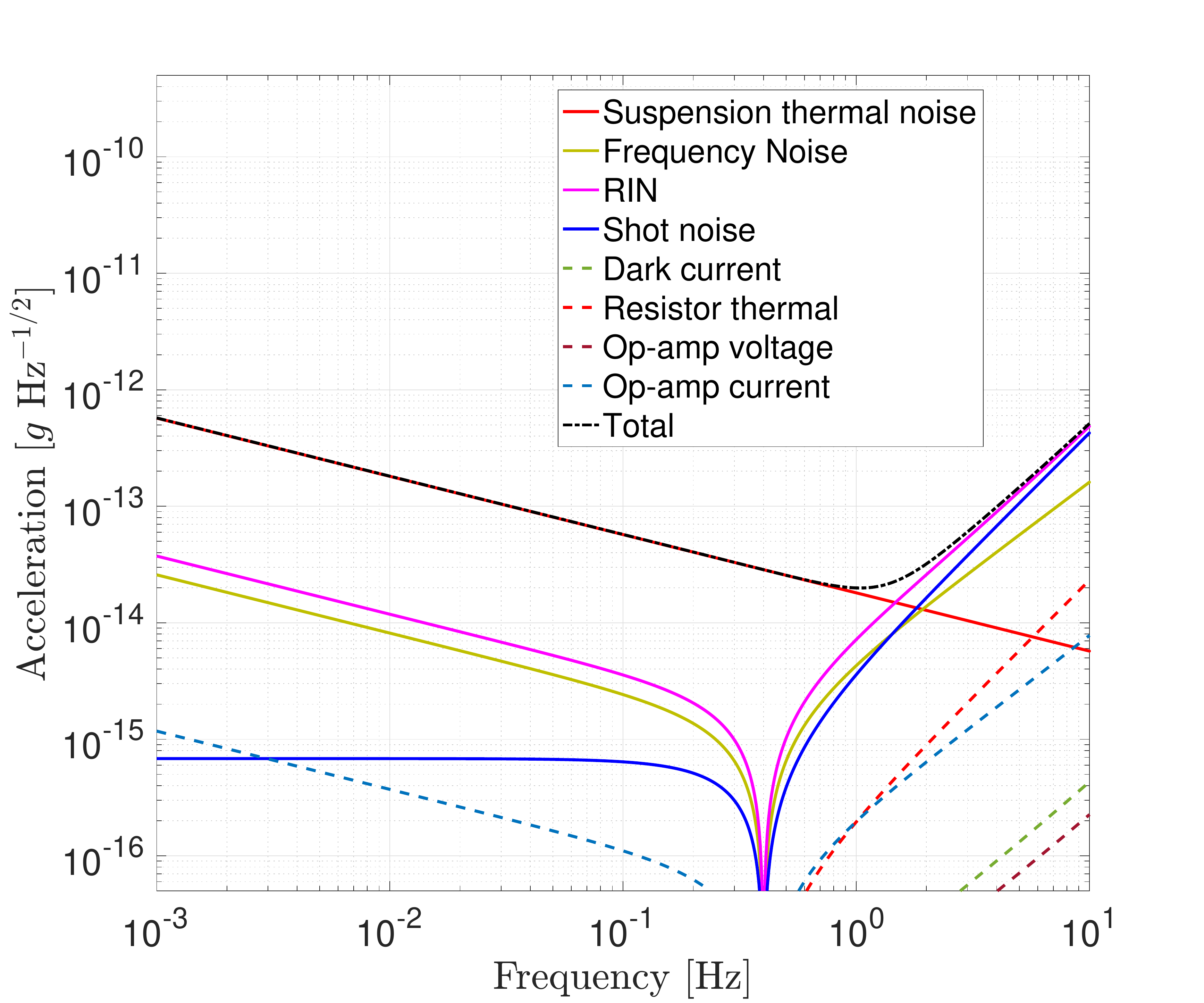}
    \label{sfig:NBmain_a}}
\caption{Minimum detectable inertial \subref{sfig:NBmain_d} displacement and \subref{sfig:NBmain_a} acceleration for a structurally damped accelerometer with interferometric readout as in Fig.~\ref{fig:SetupStart}. In this noise budget the suspension natural frequency of the accelerometer was assumed to be 0.4 Hz. The Peterson noise models are not visible as they lie above the vertical scale.}
\label{fig:NBmain}
\end{figure}

In table~\ref{tab:NBpar}, parameters similar to those used in Ref.\,\cite{IFOsens2} are presented. A higher quality factor and lower temperature sharply reduce the thermal noise contribution to the noise budget, which is shown in figure~\ref{fig:NBmain}. 

In this particular configuration, the accelerometer mechanical quality factor was found to be limited by viscous damping associated with eddy currents induced on the closely spaced moving metal surfaces by the VC stray field. Here, the aim is to be structurally damped, which will cause a thermal Brownian noise of\,\cite{SaulsonBook}  
\begin{equation}
x_{\rm{th}}^2 = \frac{4k_{\rm{B}} T k \phi}{(k-m\omega_0^2)^2+k^2\omega^2}\frac{1}{\omega} \\
\end{equation}
where $k_{\rm{B}}$ denotes the Boltzmann constant, $T$ the temperature and $\phi = 1/Q(\omega)$ the structural loss angle. With $\omega$ the angular frequency of the input vibration and $k$ the stiffness of the oscillator under study, $\omega_0$ denotes the natural frequency of the suspension. It can be seen that the displacement amplitude spectral density (ASD) $x_{\rm{th}} \propto \omega^{-2.5}$ above the resonance frequency. 

Below, calculation methods for several noise sources are summarized from ref.\,\cite{IFOsens2}. The shot noise limit can be calculated to be
\begin{equation}
	i_{\rm{sn}} = \sqrt{2 e I_{\rm{PD}}} = \sqrt{2e\rho P_{\rm{PD}}}, 
\end{equation} 
where $e$ denotes the elementary charge and $\rho$ the responsivity in A/W of the photodiode. 

For solid state lasers the Relative intensity noise (RIN) spectrum can be roughly expressed as
\begin{equation}
i_{\rm{RIN}} = i_{\rm{sn}} \sqrt{\frac{\omega_{\rm{c}}}{\omega}+1},
\end{equation}
where $\omega_{\rm{c}}$ represents the corner frequency above which the light source intensity fluctuations converge to shot noise limit. Thanks to the differential configuration of the interferometer $\omega_{\rm{c}}$ can be pushed to low frequency. The effective value of $\omega_{\rm{c}}$ can be determined experimentally. In ref.\,\cite{MyThesis} the used differential amplifier is able to get  $\omega_{\rm{c}}$ down to about 5 Hz.

Laser frequency noise can also impact the total noise budget since a frequency noise $\nu_{\rm{L}}$ (in Hz/$\sqrt{\rm{Hz}}$) translates into a readout displacement noise 
\begin{equation}\label{eq:FreqNoise}
x_{\rm{f}} = \frac{\nu_{\rm{L}}}{\nu_{\rm{0}} } \Delta L_0,
\end{equation} 
where $\nu_{\rm{L}}$ represents the frequency noise quoted by the laser manufacturer, $\nu_{\rm{0}} = c / \lambda$ the central frequency and $\Delta L_0$ the static arm length difference. 

Depending on the quality of the control and readout electronics, the dynamic range can, at time of writing, be extended by approximately 8 orders of magnitude. This is important to obtain fm\,Hz$^{-1/2}$ sensitivity at high frequency and still have almost a $\mu$m range to cope with residual motions of the stage on which the sensor is mounted. Low frequency dynamic range may be impeded somewhat by the effective dynamic range of electronics being reduced because of \textit{e.g.} flicker noise. 

\begin{table} [h]
  \centering
 \begin{tabular}{  l  c  c  }
  \hline
   Parameter & Value & Unit \\
 \hline
  \hline
  Proof mass & 0.85 & kg \\
  Leg mass & 80 & g  \\
  Leg length & 7.1 & cm \\
  Natural frequency & 0.4 & Hz \\
  Quality factor & 1$\cdot 10^{4}$ & - \\
  Frequency noise\,\cite{RockSpecs}  & 500 $\cdot~f^{-1/2}$ & Hz Hz$^{-1/2}$ \\
  Static differential arm length & 0.5 & mm \\  
  Injected power & 50 & mW \\
  Wavelength & 1550 & nm \\
  Temperature & $<$ 9.2 & K \\
  Opamp voltage noise @ 100 Hz & 4.0 & nV Hz$^{-1/2}$ \\
  Opamp voltage noise @ 0.1 Hz & 50 & nV Hz$^{-1/2}$ \\
  Opamp current noise @ 100 Hz & 2.2 & fA Hz$^{-1/2}$ \\
  Feedback resistor & 20 & k$\Omega$ \\
  Diode responsivity & 1.0 & A/W \\
  Diode dark current & 50 & nA \\
  Actuator gap & 0.1 & mm \\
  \hline 
\end{tabular}
 \caption[Optomechanical and readout electronics parameters for the prototype accelerometer]{Optomechanical and readout electronics parameters for the prototype accelerometer. The modeled laser source is The Rock$^{\rm{TM}}$ from NP Photonics, the opamp used in the transimpendance amplifier is the OPA827 and the photodiodes have a typical responsivity and dark current. Some quoted electronical noise figures are at room temperature and might improve.}\label{tab:NBpar} 
\end{table}

\section{Possible applications}\label{pa}
As some future gravitational wave detectors designs involve cryogenics, these sensors could be installed and used as monitoring or an error signal generating channel depending on the furture suspension designs. As the test mass is already in a cryogenic environment, the cryogenic infrastructure needed for this sensor to operate would already be there and the small mass would not contribute significantly to the heat load. Having sub-femtometer sensing from 5\,Hz onward at that suspension stage is of the utmost importance to reach future GW detector low frequency goals.

It could also operate as a standalone sensor as it can detect all seismic conditions on Earth with a SNR of $>10^5$ between 10\,mHz - 100\,Hz. It would require a cryostat which would make it more challenging. Additionally, any application on a future particle collider such as the International Linear Collider (ILC)\,\cite{ILC} or Future Circular Collider (FCC)\,\cite{FCC} could be interesting as cryogenics are frequently used for superconducting electromagnets. The electronics of the readout can then be moved elsewhere by use of fibers as already presented and proven in the appendix of ref.\,\cite{MyThesis}.

\begin{figure}[h]
\centering
	\subfigure[~]{
	\includegraphics[width=0.47\textwidth]{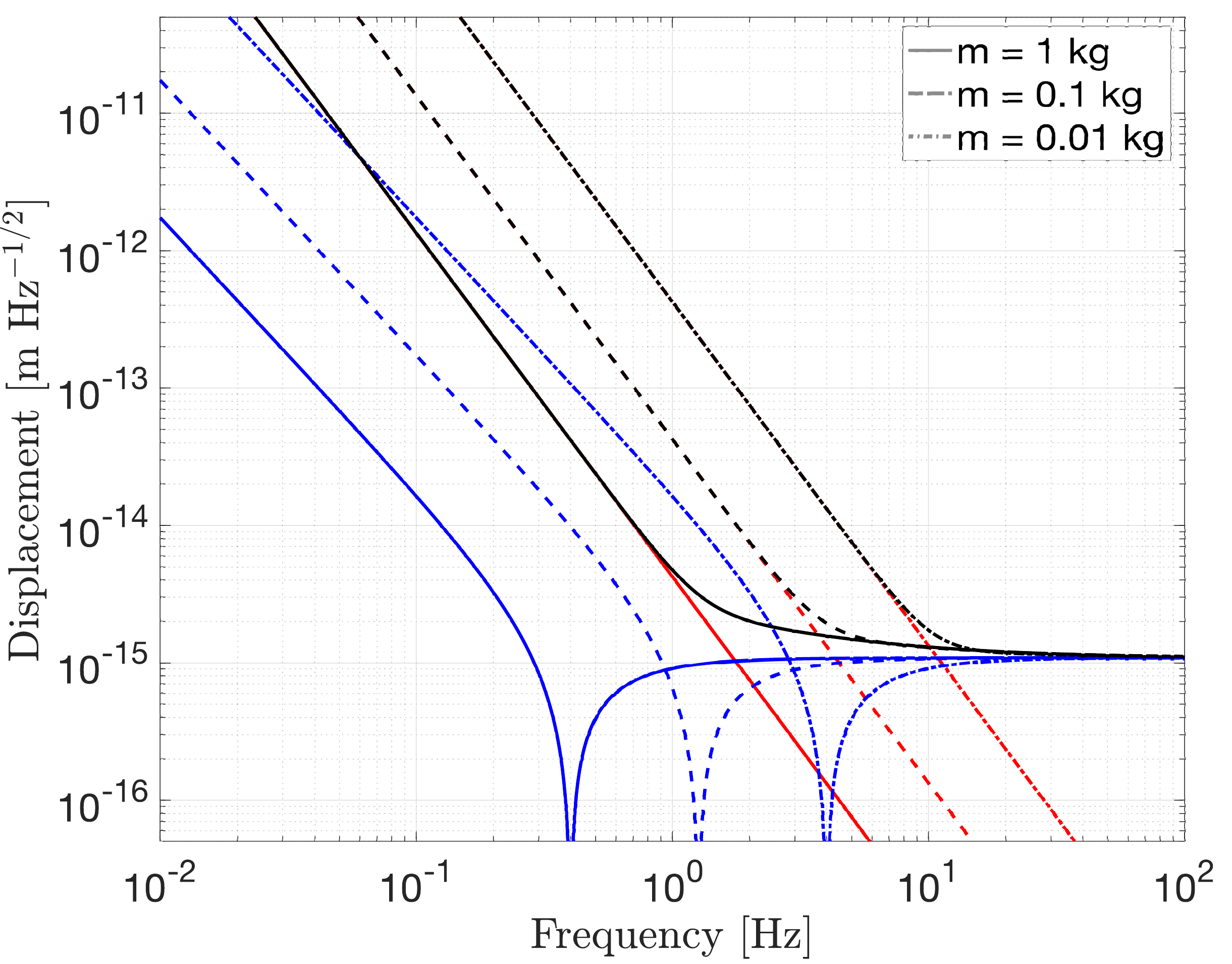}
    \label{sfig:DiffMasses_d}}
	\subfigure[~]{
	\includegraphics[width=0.47\textwidth]{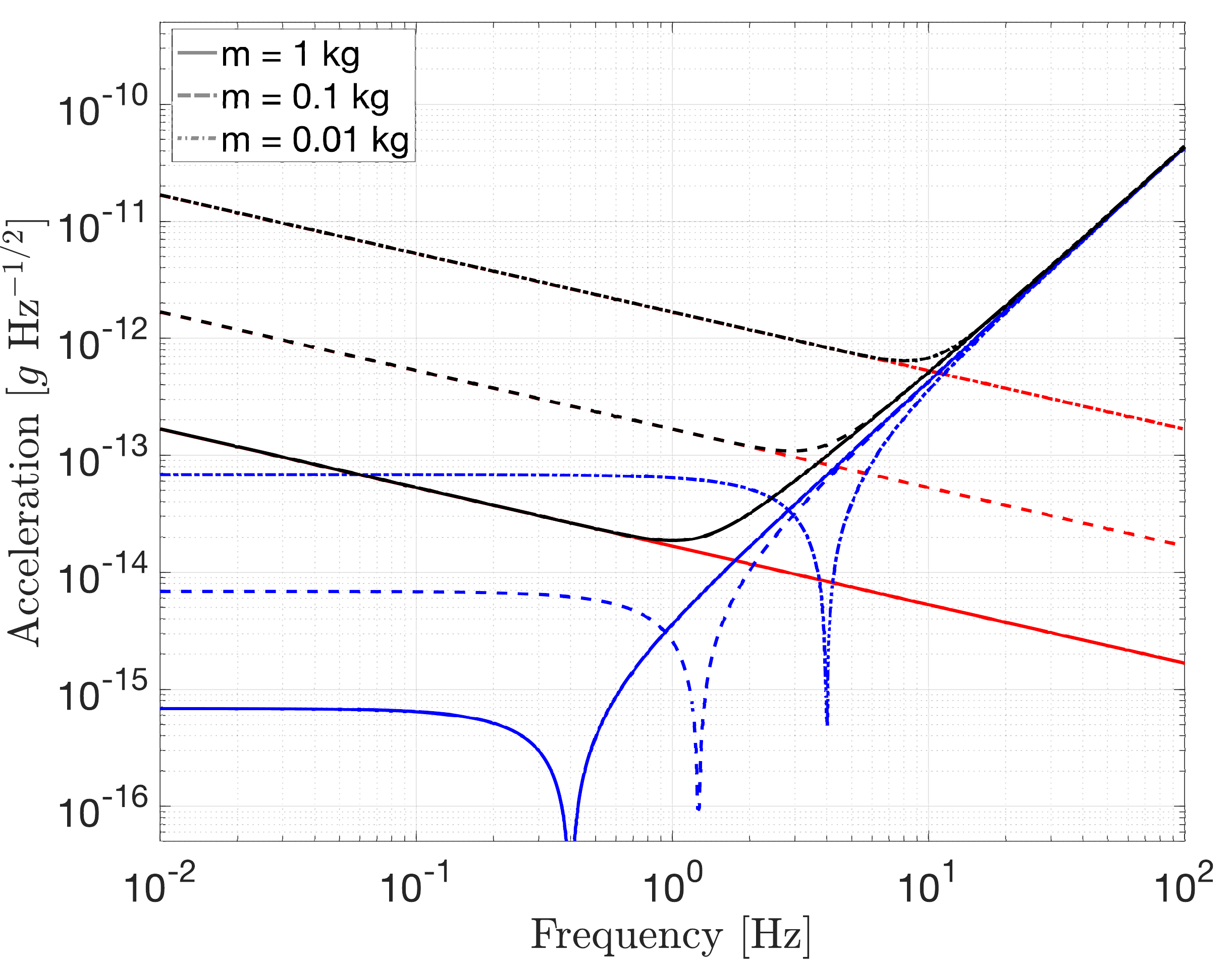}
    \label{sfig:DiffMasses_a}}
\caption{Minimum detectable inertial \subref{sfig:DiffMasses_d} displacement and \subref{sfig:DiffMasses_a} acceleration (note a difference frequency range plotted than Fig~\ref{sfig:NBmain_a}) for a structurally damped accelerometer for different proof mass values. The stiffness of the oscillator is kept constant and thus the resonance frequency goes up with $\sqrt{1/m}$. The $Q$ is kept constant at a now conservative value of $10^4$. Legend colours of Fig.~\ref{fig:NBmain} is used.}
\label{fig:DiffMasses}
\end{figure}

The analysis in previous sections focuses on adaptation of the design similar to the $\mathcal{O}$(1)\,kg proof mass published earlier~\cite{IFOsens2}. Obviously, the proof mass does not have to be order 1\,kg. In Fig.~\ref{fig:DiffMasses} the effect of changing this proof mass value is presented. The stiffness of the suspension is held constant and a conservative constant $Q$ is adopted. Note that similar sensitivity as the room-temperature 1\,kg versions is obtained by the 10\,g cryogenic version. This shows possible scaling of the sensor and the sensitivity of two other examples.

\section{Conclusion}
A novel cryogenic accelerometer that promises to reach a broadband sub-femtometer sensitivity from several hundred mHz to several hundred Hz is presented. The noise budget shows fm Hz$^{-1/2}$ sensitivity levels are possible from about 5 Hz onwards. This corresponds to a $<500$ f$g$ Hz$^{-1/2}$ acceleration sensitivity from 1\,mHz - 10\,Hz with a maximum sensitivity of 10\,f$g$\,Hz$^{-1/2}$ around 1\,Hz.

To increase dynamic range, the sensor is designed to include a feedback loop, which uses a coil magnet actuator. In prior work, this actuator decreased the $Q$ factor which was limiting suspension thermal noise. Now, by operating at cryogenic temperatures and using superconducting material, this Eddy current damping effect is eliminated. Currently, a proposed change of design of the actuator is being investigated at Nikhef. The design aims to decrease Eddy current damping by switching coil and magnet to have the magnet attached to moving parts\,\cite{PCnelson}. This results in $Q$ factors up to 6000\,\cite{JoshGWADW} at the expense of using k$\Omega$ series resistors with the coil. This would mean high voltage operation, which for GW suspension application would be challenging.

This order of magnitude improvement over earlier room temperature and non-superconducting versions of this sensor design brings about even more ability to also monitor the final stages of a GW detector. Additionally, this work will benefit precision measurements in geophysics and gravimetry as well as the use as error signal generation for vibration isolation control in particle accelerators.

\begin{acknowledgements}
The author would like to thank F. Van Kann for sharing his experience regarding Niobium spiral transducers used here as actuators. Additionally, A. Bertolini, N. de Gaay Fortman, G. Hammond, C. Blair, D.G. Blair, J. Harms, R. DeSalvo and P. Murray have contributed with valuable comments and discussions. This work is funded by the ARC Centre of Excellence for Gravitational Wave Discovery OzGrav.
\end{acknowledgements}

$~$


\begin{thebibliography}{99}
 \bibitem{IFOsuggest} M.E. Gertsenshtein and V.I. Pustovoit, "On the detection of low frequency gravitational waves", Sov. Phys. JETP 16, num. 2, pp 433-435 (1962)
	\bibitem{GWfirst} B.P. Abbott \textit{et al.}, "Observation of Gravitational Waves from a Binary Black Hole Merger", Phys. Rev. Lett., vol. 116, pp. 061102 (2016)
	\bibitem{MMgw} B.P. Abbott \textit{et al.}, "GW170817: Observation of Gravitational Waves from a Binary Neutron Star Inspiral", Phys. Rev. Lett., vol. 119, pp. 161101 (2017)
	\bibitem{MMall} B.P. Abbott \textit{et al.}, "Multi-messenger Observations of a Binary Neutron Star Merger", ApJL, 484(2), L12 (2017)
	\bibitem{LIGO} B. Barish and R. Weiss, "LIGO and the Detection of Gravitational Waves", Physics Today 52, 10, 44 (1999)
	     \bibitem{iVirgo} T. Accadia \textit{et al.}, "Performance of the Virgo interferometer longitudinal control system during the second science run", Astroparticle Physics, 34, Issue 7, pp 521-527 (2011)
		\bibitem{aLIGO} B. Abbott \textit{et al}., "LIGO: the Laser Interferometer Gravitational-Wave Observatory", Rep. Prog. Phys. 72 076901 (2009)
	  
		\bibitem{AdVirgo} T Accadia \textit{et al.}, "Virgo: a laser interferometer to detect gravitational waves", Journal of Instrumentation, 7, P03012 (2012)
	\bibitem{KAGRA} Y. Aso \textit{et al.}, "Interferometer design of the KAGRA gravitational wave detector", Phys. Rev. D, 88, 043007 (2013)
	\bibitem{LVDT} H. Tariq \textit{et al.}, "The linear variable differential transformer (LVDT) position sensor for gravitational wave interferometer low-frequency controls", NIM A, 489, pp 570-576 (2002)
	\bibitem{OSEM} K.A. Strain \textit{et al.}, Recommendation of a design for the {OSEM} sensors, LIGO note LIGO-E040108-00-K, 2004
	\bibitem{BRS} K. Venkateswara \textit{et al.}, "A high-precision mechanical absolute-rotation sensor", Rev. Sci. Instr. 85, 015005 (2014) 
	\bibitem{ALFRA} J.J. McCann \textit{et al.} "A laser walk-off sensor for high-precision low-frequency rotation measurements", Rev. Sci. Instr. 90, 045005 (2019)
	\bibitem{PLI} N. Azaryan \textit{et al.}, "The precision laser inclinometer long-term measurement in thermo-stabilized conditions (First Experimental Data)", Phys. Part. Nuclei Lett. 12: 532 (2015)
		\bibitem{L4C} R. Kirchhoff \textit{et al.}, "Huddle test measurement of a near Johnson noise limited geophone", Rev. Sci. Instr. 88, 115008 (2017)
				\bibitem{GS13Spec} Geotech Instruments LLC, "Sensors", http://www.geoinstr.com/sensors.htm
		     \bibitem{SA} G. Ballardin \textit{et al.}, "Measurement of the VIRGO superattenuator performance for seismic noise suppression", Rev. Sci. Instr. 72, 3643 (2001)
			\bibitem{Paik} J.M. Goodkind, "The superconducting gravimeter", Rev. Sci. Instr. 70, 4131 (1999)

\bibitem{NIOBE} N.P. Linthorne \textit{et al.}, "Niobium gravitational radiation antenna with superconducting parametric transducer," Physica B, 165, pp 9-10 (1990)
\bibitem{Lee} B.H. Lee \textit{et al.}, "Orthogonal Ribbons for Suspending Test Masses in Interferometric Gravitational Wave Detectors", Phys. Lett. A 339, pp 217-223 (2005)


	\bibitem{Gray} M.B. Gray \textit{et al.}, "A simple high-sensitivity interferometric position sensor for test mass control on an advanced LIGO interferometer", Opt. Quant. Electron., 31, pp 571-582 (1999)
		\bibitem{2006} A. Bertolini \textit{et al.}, "Mechanical design of a single-axis monolithic accelerometer for advanced seismic attenuation systems", Nucl. Instr. and Meth. A 556 pp 616-623, https://doi.org/10.1016/j.nima.2005.10.117 (2006)
\bibitem{IFOsens1} J.V. van Heijningen \textit{et al.}, "Interferometric readout of a monolithic accelerometer, towards the fm/$\sqrt{\rm{ Hz}}$ resolution", NIM A 824, pp 665–669 (2016)
\bibitem{Peterson} J. Peterson, "Observations and modeling of seismic background noise", U.S. Department of Interior Geological Survey (1993)
\bibitem{PCKann} Personal communication, F. van Kann, Aug 2019.
\bibitem{PresKann} F. van Kann, "VK1 - A next generation Airborne Gravity Gradiometer", Airborne Gravity 2016 Workshop (Aug 2016)
\bibitem{DBcomm} Personal communication, D.G. Blair, Aug 2019.
\bibitem{EHcomm} Personal communication, E. Hennes, Feb 2019.
\bibitem{EarthMagn} D.A.E Vares and M.A. Persinger, "Earth’s Diminishing Magnetic Dipole Moment Is Driving Global Carbon Dioxide Levels and Global Warming", International Journal of Geosciences, Vol. 6, pp. 846-852 (2015) 
\bibitem{IFOsens2} J.V. van Heijningen \textit{et al.}, "A novel interferometrically read out inertial sensor for future gravitational wave detectors", IEEE SAS proceedings, pp 76-80 (2018)
\bibitem{MyThesis} J.V. van Heijningen, "Low-frequency performance improvement of seismic attenuation systems and vibration sensors for next generation gravitational wave detectors", Ph.D. thesis, VU University Amsterdam (2018)	
\bibitem{Bertolini} A. Bertolini \textit{et al.}, "Mechanical design of a single-axis monolithic accelerometer for advanced seismic attenuation systems", Nucl. Instr. and Meth. A, vol. 556, pp. 616-623 (2006)
\bibitem{wee-g} B.A. Boom \textit{et al.}, "Nano-g Accelerometer using geometric anti-springs", 2017 IEEE 30th International Conference on MEMS proceedings, pp 33-36 (2017)
\bibitem{MEMS} R.P. Middlemiss \textit{et al.}, "Measurement of the Earth tides with a MEMS gravimeter",  Nature 531(7596), pp 614-617 (2016)
\bibitem{LNDeq} F. London and H. London, "The Electromagnetic Equations of the Supraconductor", Proc. of the Royal Society 149 (866): 71 (1935)
	\bibitem{Callen} H.B. Callen and T.A. Welton, "Irreversibility and Generalized Noise", Phys. Rev. 83, 34 (1951)
	\bibitem{SaulsonBook} P. Saulson, "Fundamentals of interferometric gravitational wave detectors", World Scientific Singapore, 1st ed. (1994)
		\bibitem{RockSpecs} http://www.npphotonics.com/images/pdfs/products/9
	\_NPP\_RockMod.pdf, viewed 31/01/2014 
	\bibitem{ILC} J. Brau \textit{et al.}, "ILC Reference Design Report Volume 1 - Executive Summary", arXiv:0712.1950 (2013)
	\bibitem{FCC} CERN collaboration, "The FCC Conceptual design report", fcc-cdr.web.cern.ch (2019)
	\bibitem{PCnelson} Personal communication A. Bertolini and N. de Gaay Fortman, May 2019.
	\bibitem{JoshGWADW} J.J. McCann \textit{et al.}, "Low frequency precision sensor experiments at UWA", presentation GWADW, May 2019




\end{thebibliography}
\end{document}